# Programmable multipartite entanglement analogues

Long Chen[1], Qing Chun Yin[1], Zi Xuan Cai[1], Zhou He[1], Zhi Cai Yu[1], Zhihao Lan[2], Xiaohan Cui[1], Jian Wei You[1, *], and Tie Jun Cui[1, *]

[1] State Key Laboratory of Millimeter Waves, Southeast University, Nanjing 210096, China.

[2] College of Physical Sciences and Engineering, Mohammed VI Polytechnic University, Ben Guerir, 43150, Morocco.

[*] Email: jvyou@seu.edu.cn, and tjcui@seu.edu.cn

Quantum entanglement underpins the quantum information science. Its multipartite forms distribute coherence across several subsystems, but controlling their composition and order is increasingly difficult as more subsystems participate. Classical nonseparable waves provide accessible analogues of these structures, although existing implementations largely fix the participating degrees of freedom through their generating architectures. Here we overcome this constraint by making multipartiteness itself programmable in electromagnetic fields. A spatiotemporal metasurface reconfigures the identity and number of coherently linked state spaces independently across polarization, frequency, orbital angular momentum, wavevector direction, and orthogonal envelope modes. We realize complete Bell-like and Greenberger-Horne-Zeilinger-like families from bipartite to five-partite order and exchange constituents at fixed order. To our knowledge, the five-partite state has the highest multipartite order experimentally realized in classical electromagnetic fields. These results establish the distribution of coherence across physical state spaces as a programmable property of classical waves and provide a platform for studying selected structures of multipartite quantum states without invoking quantum entanglement or nonlocality.

The structure of a composite physical state depends on the dimensions of its constituent spaces and the number of spaces across which coherence is distributed. A single degree of freedom can be high-dimensional without being multipartite, whereas the relative phase of an $N$-partite Greenberger-Horne-Zeilinger (GHZ) state stores its relative phase only in the coherence of the full tensor product. Multipartite order thus describes the physical extent of coherence across independently resolvable state spaces, distinct from the dimension of any one constituent space. This distinction is central to high-dimensional entanglement and quantum structured light[1-8]. The classical electromagnetic (EM) waves provide a setting in which this state-space hierarchy can be studied separately from genuine quantum resources. Polarization, spatial mode, frequency, wavevector and orbital angular momentum (OAM) can furnish distinct vector spaces whose coherent coupling generates fields that cannot be factorized with respect to a prescribed physical partition[9-11]. These fields remain classical and local: they imply neither quantum entanglement nor Bell nonlocality. Nevertheless, their tensor-product structure supports Bell-like and GHZ-like analogues, quantum-inspired transformations, tomography and fidelity-based characterization[12-20]. Experiments have progressed from bipartite nonseparability to complete tripartite GHZ-like families and four-partite classical states[21-24]. These advances establish that classical waves can sustain increasingly high-order correlations, while the physical constituents of these correlations remain largely fixed by the generating architecture.

Controlling multipartite structure requires independently addressable physical observables of the same field to participate coherently in a common nonfactorizable state. Enlarging a local basis or adding addressable observables alone does not increase multipartite order. An additional state space must instead enter the same phase-coherent branch structure. The transition from multidimensional to multipartite control therefore shifts the object of control from states within a prescribed product space to controlling the physical partition across which coherence is distributed. In existing classical implementations, cavity modes, resonator trajectories, propagation geometries or mode-conversion processes typically predetermine the identity and number of participating state spaces[21-24.] Extensions to five-partite classical GHZ-like structures have been proposed theoretically but, to our knowledge, have not been realized experimentally[22]. Spatiotemporal metasurfaces can overcome this architectural constraint by combining the spatial and temporal scattering in a reconfigurable aperture[25-39]. Their programmable responses provide simultaneous accesses to the polarization, spectral and spatial observables in a unified scattering platform. However, multidimensional addressability alone does not constitute multipartite control. Multipartite control instead requires branch-conditioned coherent scattering that determines which state spaces enter a common nonfactorizable relation and how many of them share the resulting global coherence.

Here we make the multipartite organization of an EM wave programmable in a single physical platform. A dual-polarization spatiotemporal metasurface independently programs the spatial and temporal responses of two orthogonal polarization channels, enabling conditional scattering across five EM state spaces associated with the polarization $P$, frequency $F$, wavevector $K$, OAM $L$ and orthogonal amplitude-envelope modes $A$. This programmability separates two independent structural variables of the composite field, its physical composition and its multipartite order. At a fixed order, exchanging the participating state spaces transfers the same Bell-like or GHZ-like correlation structure between physically distinct compositions. Increasing the order from two to five extends a common coherent branch relation across additional independently resolvable observables, expanding the associated binary product space from 4 to 32 dimensions. The hierarchy culminates in a five-partite GHZ-like state involving all five available EM state spaces participate in a single coherent relation. To our knowledge, this constitutes the highest multipartite order experimentally realized in a classical EM field. State-resolved measurements of the complete Bell-like and GHZ-like families, together with reconstructed density matrices and high fidelities, verify the prescribed composition and multipartite order throughout this hierarchy. Together, these results establish programmable multipartiteness as a distinct regime of wave control in which both the identity and the number of state spaces sharing global coherence become reconfigurable properties of the field itself.

## Programming multipartite state structure

Multipartite structure is defined by coherent correlations among distinct state spaces rather than in the individual spaces themselves. Hence, an EM field can be resolved into independently addressable degrees of freedom, each spanning a constituent vector space $\mathcal{H}_\mu$, and describe its nonseparability through the coherent relations among these spaces. The simplest realization contains two binary factors. A two-qubit Bell state and a classical Bell-like EM state share the equivalent nonfactorizable tensor-product structure, while the latter remains entirely classical and local (**Fig. 1a**). In the polarization-OAM realization (**Fig. 1b**), polarization $P$ and OAM $L$ form the two constituent state spaces, giving $|\Psi_{PL}\rangle_\mathrm{C} = \frac{|p_0\rangle|\ell_0\rangle + e^{i\phi}|p_1\rangle|\ell_1\rangle}{\sqrt{2}}$, with $\langle p_0|p_1\rangle = \langle \ell_0|\ell_1\rangle = 0$. Neither polarization nor OAM alone determines the composite state. Its defining information resides in the coherent correlation between the two factors, so that the projection onto one polarization basis state conditionally selects the corresponding OAM component according to ${}_P\langle p_i|\Psi_{PL}\rangle_\mathrm{C} \propto |\ell_i\rangle$. This conditional relation distinguishes the nonseparability from conventional multidimensional multiplexing, in which several observables coexist within the same field while remaining independently specified. Throughout this work, ket notation represents normalized classical field states in the corresponding EM vector spaces. The quantum-classical state-space correspondence and its physical limits are detailed in **Supplementary Note 1**.

Introducing an additional state-space factor changes how coherence is distributed across the composite field rather than merely enlarging the available basis. The correspondence with a tripartite GHZ state illustrates this transition (**Fig. 1c**). Its classical analogue comprises two complementary product branches that differ in every participating EM degree of freedom. One such realization combines polarization $P$, wavevector $K$ and OAM $L$ (**Fig. 1d**), with the global coherence linking $|p_0\rangle|k_0\rangle|\ell_0\rangle$ and $|p_1\rangle|k_1\rangle|\ell_1\rangle$. To express this structure independently of the observables involved, each programming configuration $q$ is associated with an active set of state spaces $\mathcal{S}_q = \{\mu_1, \ldots, \mu_{N_q}\}$. Each factor contributes an orthonormal binary basis $\{|0\rangle_{\mu_j}, |1\rangle_{\mu_j}\}$, yielding the canonical multipartite family

$$\left|\Psi_{q,\boldsymbol{b}}^{\pm}\right\rangle_{\mathrm{C}} = \frac{\left|\boldsymbol{b}\right\rangle_{\mathcal{S}_q} \pm \left|\bar{\boldsymbol{b}}\right\rangle_{\mathcal{S}_q}}{\sqrt{2}}, \qquad \left|\boldsymbol{b}\right\rangle_{\mathcal{S}_q} = \bigotimes_{j=1}^{N_q} \left|b_j\right\rangle_{\mu_j} \tag{1}$$

Here, $\boldsymbol{b}$ represents one member of each complementary pair $\{\boldsymbol{b}, \bar{\boldsymbol{b}}\}$; we choose $b_1 = 0$ throughout without loss of generality. The $2^{N_q-1}$ complementary branch pairs together with the two relative signs generate $2^{N_q}$ mutually orthogonal Bell-like or GHZ-like states. The complete binary Bell/GHZ basis construction is derived in **Supplementary Note 2**. Because the two branches are orthogonal in every participating factor, the ideal state has Schmidt rank two across every nontrivial bipartition. Increasing the order from two to three therefore extends the coherence across the full multipartite partition, producing a correlation structure that cannot be decomposed into independent lower-order correlations. The distinction between physical composition and multipartite order becomes explicit once the active factors are separated from the full set of EM state spaces available to the platform. For our platform, the available set is $\mathfrak{D} = \{P, F, K, L, A\}$, comprising polarization, frequency, wavevector, OAM and orthogonal amplitude-envelope modes. Their participation in configuration $q$ is encoded by $m_{q,\mu} \in \{0,1\}$, giving

$$\mathcal{S}_q = \{\mu \in \mathcal{D} \mid m_{q,\mu} = 1\}, \qquad N_q = \left|\mathcal{S}_q\right| = \sum_{\mu \in \mathfrak{D}} m_{q,\mu}, \qquad D_q = \prod_{\mu \in \mathcal{S}_q} d_\mu \tag{2}$$

These quantities resolve three complementary aspects of the composite-state structure. The active set $\mathcal{S}_q$ specifies the physical composition of the nonseparable state, $N_q$ defines its multipartite order through the number of coherently participating factors, and $D_q$ gives the dimension of the associated product space. For the binary state spaces used here, $d_\mu = 2$ and $D_q = 2^{N_q}$, yielding dimensions of 4, 8, 16 and 32 as the multipartite order increases from two to five (**Fig. 1e**). The resulting growth of $D_q$ is therefore a consequence of extending a common coherent relation across additional state spaces rather than the defining feature of multipartiteness itself. Increasing the local dimension of a single factor can enlarge $D_q$ without changing $N_q$. This separation between composition and order establishes the structural basis of programmable multipartiteness. The distinction among state-space composition, multipartite order and composite dimension is developed in **Supplementary Note 3**. Different active subsets of the same EM state-space set can support the same multipartite order while representing physically inequivalent composite states. The four-partite examples in **Fig. 1f** illustrate this constituent exchange, whereas the five-partite configuration incorporates all five available state spaces into a single nonfactorizable relation. Here $K$ labels distinct wavevector-direction states, while $A$ labels mutually orthogonal envelope modes. The modal definitions, vector-field representation and coherent-projection formalism for the five EM state spaces are given in **Supplementary Note 4**. The available EM degrees of freedom therefore remain fixed, while programming determines which of them enter the active nonseparable sector and how many are coherently linked within it. In this formulation, state-space composition and multipartite order emerge as independently programmable structural attributes of the same EM field.

## Interchangeable constituents of bipartite nonseparability

The distinction between multipartite order and state-space composition is already manifest in the bipartite sector. Setting $N_q = 2$ fixes the number of participating state spaces while allowing their physical identity to vary. Bell-like nonseparability can therefore be encoded in physically different tensor-product spaces with the same underlying Schmidt structure. Control over composition requires this nonfactorizable structure to remain invariant when one physical constituent is exchanged. This identifies composition as a structural variable of the correlation itself. The dual-polarization spatiotemporal metasurface implements this constituent exchange through conditional scattering within one aperture (**Fig. 2a**). The two orthogonal polarization channels share the same physical aperture while undergoing independently programmed spatial and temporal transformations. Neglecting weak cross-polarized leakage, the scattering operator is approximately block-diagonal in the polarization basis, $\widehat{\mathcal{M}}_q \simeq \sum_{\alpha=0}^{1} |p_\alpha\rangle\langle p_\alpha| \otimes \widehat{T}_{\alpha,q}$, where $q$ labels the programming configuration and $\widehat{T}_{\alpha,q}$ denotes the transformation conditioned on polarization state $|p_\alpha\rangle$. Polarization therefore acts as the control space for conditional transformations. The correspondence between conditional EM scattering and controlled state preparation is developed in **Supplementary Note 5**. The four-diode meta-atom provides independently addressable 2-bit phase responses for the two polarization channels, implementing these branch-conditioned transformations (**Fig. 2b,c**). For a common incident mode $|u_{\text{in}}\rangle$, the two polarization branches are mapped onto normalized conditional states $|u_{0,q}\rangle$ and $|u_{1,q}\rangle$ of the selected second state-space factor. The resulting bipartite field is

$$|\Psi_q\rangle_{\text{C}} = \frac{a_{0,q}|p_0\rangle|u_{0,q}\rangle + e^{i\phi_q} a_{1,q}|p_1\rangle|u_{1,q}\rangle}{\sqrt{|a_{0,q}|^2 + |a_{1,q}|^2}} \tag{3}$$

where $a_{0,q}$ and $a_{1,q}$ are real non-negative branch amplitudes and $\phi_q$ is their relative phase. The resulting nonseparability is determined by the branch balance and the mutual overlap of the two conditional states. The normalized branch population is $p_q = \frac{|a_{0,q}|^2}{|a_{0,q}|^2+|a_{1,q}|^2}$, while $\gamma_q = \langle u_{1,q}|u_{0,q}\rangle$ quantifies the overlap of the conditional states. The eigenvalues of the reduced polarization density matrix, equivalently the squared Schmidt coefficients, are

$$\lambda_{\pm}^{(q)} = \frac{1}{2}\left[1 \pm \sqrt{1 - 4p_q(1-p_q)\left(1 - |\gamma_q|^2\right)}\right] \tag{4}$$

The maximally nonseparable Bell-like limit is obtained for balanced branches, $p_q = 1/2$, and orthogonal conditional states, $\gamma_q = 0$, for which $\lambda_+ = \lambda_- = 1/2$. Conversely, the state factorizes when one branch vanishes, $p_q = 0$ or 1, or when the two conditional states coincide up to a phase, $|\gamma_q| = 1$. Crucially, for fixed $p_q$ and $|\gamma_q|$, the Schmidt spectrum is independent of the physical observable used for the second binary factor. It is determined solely by the branch balance and conditional-state overlap. The full Schmidt analysis of programmable bipartite nonseparability is provided in **Supplementary Note 6**. This invariance of the Schmidt structure is realized in four physically inequivalent state-space compositions (**Fig. 2d-g**). All the active sets $\mathcal{S}_q = \{P, K\}$, $\{\{P, L\}$, $\{P, F\}$ and $\{P, A\}$ have $N_q = 2$, while the second tensor factor is carried by a different EM observable in each case. A spatial phase gradient maps the polarization branches onto distinct wavevectors for $P - K$, whereas azimuthal phase winding generates orthogonal OAM states for $P - L$. Temporal modulation supplies discrete frequency states for $P - F$, while orthogonal field envelopes define the $A$ basis of $P - A$. Although these transformations act in different physical domains, they are programmed to reproduce the same two-branch Bell-like Schmidt structure. Changing $\mathcal{S}_q$ thus exchanges the physical tensor factor itself rather than merely relabeling basis modes within a fixed degree of freedom.

Experimental field measurements are then performed to test whether this Schmidt structure is preserved across physically distinct compositions. Projection onto each four-dimensional product basis yields normalized complex modal coefficients, used to reconstruct the corresponding projected-state density matrix. The real and imaginary components in **Fig. 2h-k** resolve the expected branch populations and interbranch coherence for representative $\Psi_0^+$ states. The complete Bell-like families provide a more stringent test because both complementary basis pairings and their relative phases must be preserved. Across all sixteen combinations of Bell-like state and physical composition, the projected-state fidelities range from 94.9% to 99.0% (**Fig. 2l-o**), evaluated as $\mathcal{F} = \langle \Psi_{\text{tar}} | \rho_{\text{meas}} | \Psi_{\text{tar}} \rangle$. The ideal Schmidt structure is common to all four compositions, while the reconstructed spectra quantify composition-dependent departures from that reference (**Supplementary Notes 6** and **7**). The exchanged observables therefore support the same target correlation structure through physically different field modes. The physical constituents of nonseparability can therefore be exchanged while the underlying correlation structure is retained, establishing state-space composition as a programmable structural coordinate of the EM wave. Extended characterization of the complete Bell-like families is provided in **Supplementary Note 7**.

## Global coherence in tripartite states

The bipartite results establish state-space composition as an independent structural variable at fixed multipartite order. Increasing the multipartite order to $N_q = 3$ changes not only the number of participating factors but also how coherence is encoded across them. Three binary factors span an eight-dimensional product space, yet the essential change lies not in the doubling of dimension but in the emergence of coherence that belongs to the full three-factor state. For the canonical GHZ-like family, the relative phase resides in the off-diagonal coherence between two complementary three-factor branches. Tracing out any factor removes this phase information from the remaining two-factor state. The relative phase is therefore absent from every two-factor reduced state and is retained in the coherence of the complete three-factor field. Increasing multipartite order therefore redistributes coherence across the physical partition of the field rather than merely enlarging its accessible basis.

The same programmable aperture realizes this tripartite coherence through joint spatial and temporal control of the two polarization branches (**Fig. 3a**). Each configuration selects three binary factors from the available set $\mathfrak{D} = \{P, F, K, L, A\}$, fixing $N_q = 3$ and $D_q = 8$ while allowing the physical composition to vary (**Fig. 3b**). At this fixed order and dimension, the participation map in **Fig. 3c** identifies four physically distinct compositions, $\{P, K, L\}$, $\{P, F, A\}$, $\{P, F, K\}$ and $\{P, F, L\}$. All four configurations share $N_q = 3$ and $D_q = 8$, while the physical observables carrying the coherence are different. This isolates composition from both multipartite order and dimensionality, allowing the physical carriers of the correlation to change without altering its three-factor organization. The composite-space dimension is therefore preserved while the physical factors supporting the coherence are reconfigured. The global character of the coherence becomes explicit upon tracing out any participating factor. For the canonical states of **Eq. 1**, tracing over state space $\mu_k$ yields

$$\rho_{\bar{k}}^{(\pm)} = \mathrm{Tr}_{\mu_k}\left[\left|\Psi_{q,\boldsymbol{b}}^{\pm}\right\rangle\left\langle\Psi_{q,\boldsymbol{b}}^{\pm}\right|\right] = \frac{1}{2}\left(\left|\boldsymbol{b}_{\bar{k}}\right\rangle\left\langle\boldsymbol{b}_{\bar{k}}\right| + \left|\bar{\boldsymbol{b}}_{\bar{k}}\right\rangle\left\langle\bar{\boldsymbol{b}}_{\bar{k}}\right|\right) \tag{5}$$

where $\boldsymbol{b}_{\bar{k}}$ denotes the two-factor basis string remaining after the $k$th constituent is traced out. Orthogonality of the local binary bases eliminates the off-diagonal terms linking the complementary branches, giving $\rho_{\bar{k}}^{(+)} = \rho_{\bar{k}}^{(-)}$. In this ideal orthogonal-basis limit, the relative sign cannot be recovered from any two-factor reduced state and survives only in the coherence of the complete tripartite field. The disappearance of these coherences under every partial trace provides a direct state-space signature that the relative phase is stored globally rather than in any pairwise sector. This loss of phase information under any partial trace distinguishes global tripartite coherence from a collection of independent pairwise correlations. The reduced-state origin of this global GHZ-like phase storage is derived in **Supplementary Note 8**. The $P - K - L$ composition provides a direct realization of this global three-factor structure (**Fig. 3d**). Polarization, wavevector and OAM form independently resolvable state spaces whose local basis states are correlated into two complementary three-factor branches. The reconstructed projected-state density matrix resolves their populations and the off-diagonal coherence linking the complementary states (**Fig. 3e**). These off-diagonal elements provide the experimentally accessible signature of the coherent phase relation that distinguishes the GHZ-like state from an incoherent mixture with the same branches. The element-wise deviation map in **Fig. 3f** quantifies residual departures from the ideal projected state, while the complete eight-state family in **Fig. 3g** extends the characterization to all four complementary basis pairings and both relative signs.

Crucially, the global three-factor coherence is preserved when the physical composition is reconfigured. The $P - F - A$ composition is realized in **Fig. 3h**, while replacing $A$ with wavevector $K$ or OAM $L$ yields the $P - F - K$ and $P - F - L$ states shown in **Fig. 3i** and **Fig. 3j**, respectively. Although the constituent observables and local bases differ, the reconstructed states retain the characteristic complementary-branch structure of the GHZ-like family. The relative phase remains encoded collectively across all three factors and cannot be recovered from any pairwise reduced state. The tripartite states therefore preserve their global GHZ-like coherence under constituent exchange, establishing state-space composition as programmable independently of both composite dimension and multipartite order. Complete-family verification of these tripartite states is presented in **Supplementary Note 9**.

## Scaling global coherence to five-partite order

Once coherence is distributed globally across several state spaces, increasing multipartite order requires an additional degree of freedom to enter the same nonfactorizable branch structure rather than merely enlarging the product space. The new factor must occupy different local states in the two complementary branches and thereby become an active carrier of global coherence. To express this condition independently of a particular programming configuration, let $\mathcal{S}$ denote an active $N$-partite set and let $v \notin \mathcal{S}$ denote an additional binary state-space factor. Raising the multipartite order from $N$ to $N+1$ then requires

$$|\Psi^{\pm}_{\mathcal{S}\cup\{v\},\boldsymbol{b}}\rangle_{\mathrm{C}} = \frac{|\boldsymbol{b}\rangle_{\mathcal{S}}|0\rangle_v \pm |\bar{\boldsymbol{b}}\rangle_{\mathcal{S}}|1\rangle_v}{\sqrt{2}} \tag{6}$$

Because $|\boldsymbol{b}\rangle_{\mathcal{S}}$ and $|\bar{\boldsymbol{b}}\rangle_{\mathcal{S}}$ are orthogonal, and $|0\rangle_v$ and $|1\rangle_v$ form an orthogonal local basis, the state in **Eq. 6** has Schmidt rank two across the bipartition $\mathcal{S} \mid v$. It therefore cannot be factorized as $|\Phi\rangle_{\mathcal{S}} \otimes |\chi\rangle_v$ for any states $|\Phi\rangle_{\mathcal{S}}$ and $|\chi\rangle_v$. The additional factor is therefore incorporated in the active nonseparable sector rather than appended as a separable spectator. This branch-conditioned participation increases the multipartite order without changing the local dimensionality of the pre-existing constituents. For the canonical GHZ-like family considered here, the global relative phase remains absent from every proper reduced state, while the coherence connects complementary product branches that differ across a larger number of local factors. Increasing multipartite order consequently extends the physical support of the same global coherent relation across an additional independently resolvable state space. The recursive extension to arbitrary multipartite order is derived in **Supplementary Note 10**. In the spatiotemporal metasurface, different configurations $q$ realize this scaling by selecting active subsets $\mathcal{S}_q \subseteq \mathcal{D} = \{P, F, K, L, A\}$ with different multipartite orders $N_q = |\mathcal{S}_q|$ (Fig. 4a). The available EM state spaces remain unchanged, while programming controls how many of them participate coherently in the active nonseparable sector. Four active binary factors span a 16-dimensional composite space, whereas participation of all five produces a 32-dimensional space (Fig. 4b). Although $D_q = 2^{N_q}$, the essential progression is the extension of one global coherent relation across four and ultimately five independently resolvable physical observables. In this sense, increasing $N_q$ changes the physical extent of the coherence rather than simply the number of basis states available to the field.

At $N_q = 4$, multipartite order can be held fixed while the physical composition is varied. The $P-F-K-A$ state combines polarization, frequency, wavevector and an orthogonal amplitude-envelope mode (**Fig. 4c**). The alternative $P-F-L-A$ and $P-F-K-L$ compositions are realized in **Fig. 4d,e**, respectively. All three configurations share $N_q = 4$ and $D_q = 16$, while one physical constituent is exchanged without altering the four-factor GHZ-like branch structure. Their equivalence tests whether four-partite coherence survives constituent exchange. The corresponding deviation maps quantify element-wise departures from the ideal states in the full 16-dimensional product bases. The fidelities in **Fig. 4f-h** extend this test to the complete 16-state GHZ-like family for each composition, covering every complementary branch pairing and both relative signs.

The hierarchy culminates when all five available state spaces participate in the same coherent relation. For the $P - F - K - L - A$ state, $\mathcal{S}_q = \mathfrak{D}$, $N_q = 5$ and $D_q = 32$ (**Fig. 4i**). The active nonseparable sector therefore spans the full EM state-space set of the platform. The real and imaginary parts of the reconstructed projected-state density matrix resolve the complementary five-factor branch populations together with the off-diagonal coherence connecting them across the full product space (**Fig. 4j,k**). These off-diagonal terms distinguish coherent five-factor superposition from an incoherent mixture containing the same branch populations. The element-wise deviation from the ideal target is quantified in **Fig. 4l**, while the complete 32-state family in **Fig. 4m** tests all canonical complementary branch pairings and relative phases of the five-partite basis. The progression from three to four and five factors therefore represents more than an increase in composite-space dimension. Each increase in $N_q$ extends the same coherent relation across an additional independently addressable physical state space. At four-partite order, composition remains programmable without disrupting the global GHZ-like coherence. At five-partite order, the active correlation spans all EM state spaces available to the platform. Programmable multipartiteness is therefore realized when both the identity and the number of state spaces participating in the same global coherence become independently controllable properties of the EM field. Complete four- and five-partite verification is provided in **Supplementary Note 11**.

## Discussion

Our work establishes multipartiteness as a reconfigurable physical structure of an EM field rather than a fixed property inherited from the architecture that generates it. The essential advance is the separation of two aspects of composite-state organization that are ordinarily inseparable in wave systems. State-space composition specifies which physical degrees of freedom participate in a nonseparable correlation, whereas multipartite order determines how far the same coherent relation extends across independently resolvable state spaces. The spatiotemporal metasurface renders both quantities accessible within a single wave system by coherently reorganizing the underlying EM state spaces. The resulting hierarchy therefore represents more than an increase in Hilbert-space dimension. It demonstrates that the physical support of coherence itself can be reconfigured and progressively extended, culminating in a five-partite GHZ-like structure in which all available EM state spaces participate in one global correlation. In this sense, the metasurface acts not merely on the state carried by the wave, but on the multipartite organization through which that state is constituted.

This perspective opens a broader regime of composite-state physics beyond the Bell-like and GHZ-like families considered here. Enlarging the local state spaces would allow dimensionality and multipartite order to be varied independently, while more general coherent transformations could access W-like, graph-like and other multipartite structures with qualitatively different correlation and reduced-state properties. Higher-dimensional and alternative multipartite structures are discussed in **Supplementary Note 12**. Spatiotemporal programmability further raises the possibility of treating multipartite organization as a dynamical quantity, enabling controlled evolution between distinct physical partitions and correlation orders within the same field. Such extensions shift the emphasis of structured-wave physics from synthesizing the increasingly elaborate field distributions towards controlling how coherence, information and correlations are organized across a composite physical system.

More fundamentally, programmable multipartiteness provides a controllable classical setting in which structures central to quantum mechanics can be isolated from genuinely quantum resources. Classical EM waves do not reproduce quantum nonlocality or other genuinely quantum resources, yet they can faithfully reproduce selected tensor-product, coherence and reduced-state structures associated with multipartite quantum states. This separation distinguishes structures arising from the geometry and coherence of composite Hilbert spaces from resources that require quantumness. It also separates the local mixedness created by unresolved modal labels from statistical dephasing of the complete field, as developed in **Supplementary Note 13**. It also provides an experimentally accessible arena for developing and testing concepts related to multipartite state preparation, correlation encoding and state-space geometry before their realization in quantum platforms. By making the organization of coherence itself programmable, our work places spatiotemporal electromagnetism in a new role as an analogue laboratory for composite-state physics and lays the groundwork for exploring increasingly complex structures at the interface of classical waves, quantum foundations and quantum information science.

## Methods

### Dual-polarization programmable metasurface

The experiments were performed with a reflective spatiotemporal metasurface designed to provide independent digital control of two orthogonal linear-polarization channels within the same physical aperture. The metasurface comprised $32 \times 32$ individually addressable meta-atoms with a lattice period of 30 mm, corresponding to an active aperture of $960 \times 960\ \text{mm}^2$. Each meta-atom contained four PIN diodes (SMP1345-040LF) distributed along two orthogonal current paths. Two diodes primarily controlled the scattering response of the $x$-polarized component and the remaining two controlled that of the $y$-polarized component. Each diode pair therefore provided four electronically selectable responses, giving independent 2-bit control for the two polarization channels. The instantaneous electronic state of one meta-atom was denoted by the four-bit word $[x_1 x_2 y_1 y_2]$, where $[x_1 x_2]$ and $[y_1 y_2]$ addressed the $x$- and $y$-polarized responses, respectively. The 16 joint electronic configurations consequently defined the local scattering alphabet from which the spatial and temporal coding distributions were constructed. This architecture is essential to the multipartite-state implementation because the two polarization components occupy a common aperture while remaining independently programmable. They can therefore serve as the two coherent branches of the conditional transformation described in the main text, rather than as two independently generated beams.

The meta-atoms were implemented using multilayer printed-circuit-board architecture. Patterned radio-frequency metallic layers were separated from a continuous metallic ground plane and a dedicated direct-current bias-routing network by F4B dielectric substrates with relative permittivity $\varepsilon_r = 2.65$ and loss tangent $\tan\delta = 0.001$. The dielectric layers had thicknesses of $0.254$ mm, $4$ mm and 0.508 mm. Metallized vias provided electrical connection to the active components while maintaining separation between the radio-frequency scattering structure and the bias circuitry. After fabrication of the passive multilayer structure, the four PIN diodes were mounted by surface-mount assembly. A digital controller supplied the prescribed four-bit states to the full aperture through multichannel driving electronics. For spatially programmed measurements, one coding matrix was maintained throughout the acquisition. For spatiotemporal operation, successive coding matrices were periodically updated according to the designed sequence. The same hardware was therefore used throughout the work; different multipartite configurations were realized by changing only the digitally prescribed space-time coding.

### EM calibration of the digital scattering states

The local scattering alphabet was calibrated before synthesis of the multipartite states. Full-wave EM simulations were performed in CST Microwave Studio 2022. A single meta-atom was first analyzed under periodic boundary conditions in the two transverse directions and open boundaries normal to the surface. The $x$- and $y$-polarized incidences were simulated independently, and the frequency-dependent response of the PIN diodes was retained in the EM model. For each polarization channel, the four 2-bit electronic states produced a calibrated set of complex reflection coefficients rather than an idealized phase-only alphabet. All 16 joint four-bit configurations were additionally characterized through their complex polarization response so that residual amplitude variation and weak polarization conversion could be retained when required. This calibration is important for the multipartite-state synthesis because the nonseparable structure is determined by relative complex amplitudes between branches and therefore cannot in general be inferred from phase values alone. The resulting complex-response library was used consistently in state synthesis and full-wave simulation. Finite-aperture simulations reproduced the complete $32 \times 32$ metasurface. Open boundaries were applied around the computational domain. The calculated complex fields were sampled at the same spatial planes and frequencies as the corresponding experiments.

### Programming the physical composition of the multipartite state

Each configuration $q$ was constructed directly from the state-space representation introduced in **Eqs. 1** and **2**. The target active set $\mathcal{S}_q \subseteq \mathcal{D} = \{P, F, K, L, A\}$ specified which physical observables entered the nonseparable sector, while $N_q = | \mathcal{S}_q |$ specified the multipartite order. No additional state-space definition was introduced in the numerical implementation. Polarization $P$ provided the common conditional degree of freedom. The two orthogonal polarization states defined the two branches of the programmed transformation, and different complex transformations were assigned to these branches through the independently addressable $x$- and $y$-polarization channels. The remaining state spaces were implemented according to their physical basis. The two $K$ states were distinct wavevector-direction states generated by different transverse phase gradients. The $L$ states were defined by orthogonal OAM modes with different azimuthal phase winding. The $F$ basis consisted of selected Fourier harmonics generated by periodic temporal modulation. The $A$ basis consisted of mutually orthogonal spatial-envelope modes and therefore represented a genuine modal degree of freedom rather than two scalar field amplitudes.

The spatial coding patterns were obtained by sampling the required branch-conditioned complex distributions on the physical meta-atom lattice and mapping them onto the calibrated digital scattering states. For configurations containing several spatial factors, the corresponding phase or envelope transformations were combined coherently before digital quantization. The coding operation therefore acted on the complete complex branch field rather than independently assigning labels to individual degrees of freedom. The same procedure was used from the bipartite to the five-partite sector. Increasing $N_q$ did not modify the local dimensionality of an already active factor; it introduced an additional independently resolvable state space into the common complementary-branch relation defined by **Eq. 1**. Conversely, changing $\mathcal{S}_q$ at fixed $N_q$ exchanged a physical constituent while leaving the multipartite order unchanged.

## Temporal synthesis and harmonic-state generation

Periodic temporal coding was used whenever frequency $F$ participated in the active state space. A modulation period $T$ was divided into $N_t$ equal temporal slots, where $N_t$ denotes the coding-sequence length and is distinct from the multipartite order $N_q$. If $r_{p,n}$ is the calibrated complex reflection coefficient of polarization channel $p \in \{x, y\}$ during the $n$-th slot, the complex coefficient at Fourier order $m$ is

$$r_p^{(m)} = \frac{\eta_m}{N_t} \sum_{n=0}^{N_t - 1} r_{p,n} \exp\left(+i \frac{2\pi mn}{N_t}\right), \tag{7}$$

where $\eta_m = \text{sinc}\left(\frac{m}{N_t}\right) \exp\left(+i \frac{\pi m}{N_t}\right)$ accounts for the finite duration of each temporal slot. Here $\text{sinc}(x) = \sin(\pi x)/(\pi x)$, with $\text{sinc}(0) = 1$, and the $n$-th coding state occupies $nT/N_t \leq t < (n+1)T/N_t$. The $m$-th component occurs at $f_m = f_c + m f_s$, where $f_c$ is the incident carrier frequency and $f_s = 1/T$ is the modulation frequency.

**Eq. 7** was used as the spectral representation of the temporal codes throughout this work. Because the two polarization channels were modulated independently, they acquired different complex Fourier coefficients at a given harmonic. Their relative amplitude and phase could therefore be programmed jointly with the harmonic index itself. In configurations containing $F$, frequency was consequently not treated merely as a collection of generated sidebands. The selected harmonic constituted one factor of the tensor-product state and was coherently correlated with the other participating observables. The temporal sequences were constructed from the experimentally available four-state scattering dictionaries. Candidate sequences were evaluated using their complete complex Fourier responses, and coding solutions were selected according to the overlap with the prescribed complementary-branch state. Importantly, the synthesis used the calibrated amplitude and phase of each digital state rather than assuming four ideal unit-amplitude phase values.

**Time-domain numerical simulations**

Static frequency-domain calculations cannot reproduce the EM transients generated while the active elements are switched in time. Dynamic simulations were therefore performed using a time-varying lumped-element finite-integration formulation derived from the TVLE-TDFIT framework developed for practical space-time metasurfaces[40]. In the present work, the numerical formulation was adapted to the four-diode dual-polarization meta-atom and to the experimentally applied four-bit coding sequences. The EM domain was discretized using mutually dual primal and dual grids. Integral electric and magnetic variables were assigned to the corresponding grid edges, and Maxwell's curl equations were advanced using staggered time integration. In compact matrix form, the electric-field update was written as

$$\boldsymbol{e}^{n+1} = \boldsymbol{A}_{\mathrm{e}}^{n}\boldsymbol{e}^{n} + \boldsymbol{B}_{\mathrm{e}}^{n+1/2}\left(\tilde{\mathbf{C}}\boldsymbol{h}^{n+1/2} - \boldsymbol{j}_{\mathrm{e}}^{n+1/2}\right), \tag{8}$$

whereas the magnetic-field update was

$$\boldsymbol{h}^{n+1/2} = \boldsymbol{A}_{\mathrm{h}}^{n-1/2}\boldsymbol{h}^{n-1/2} - \boldsymbol{B}_{\mathrm{h}}^{n}(\boldsymbol{C}\boldsymbol{e}^{n} - \boldsymbol{j}_{\mathrm{m}}^{n}). \tag{9}$$

Here, $\mathbf{C}$ and $\tilde{\mathbf{C}}$ denote the discrete curl operators associated with the primal and dual grids. The matrices $\mathbf{A}_{\mathrm{e},h}$ and $\mathbf{B}_{\mathrm{e},h}$ contain the local material and conformal geometric information, and $\mathbf{j}_{\mathrm{e},m}$ denote electric and equivalent magnetic source or loss terms when present. Temporal modulation enters **Eqs. 8** and **9** through the local EM parameters of grid regions representing the active switching elements. For an electrically lossy region with effective permittivity $\bar{\varepsilon}$ and conductivity $\bar{\sigma}(t)$, the electric update coefficients take the generic form

$$\alpha_{\mathrm{e}}^{n} = \frac{\bar{\varepsilon}^{n}/\Delta t - \bar{\sigma}^{n}/2}{\bar{\varepsilon}^{n}/\Delta t + \bar{\sigma}^{n}/2}, \qquad \beta_{\mathrm{e}}^{n+1/2} = \frac{g_{\mathrm{e}}}{\bar{\varepsilon}^{n+1/2}/\Delta t + \bar{\sigma}^{n+1/2}/2}, \tag{10}$$

where $g_{\mathrm{e}}$ contains the corresponding primal–dual conformal geometry factor. The magnetic update is written analogously as

$$\alpha_{\mathrm{h}}^{n-1/2} = \frac{\bar{\mu}^{n-1/2}/\Delta t - \bar{\kappa}^{n-1/2}/2}{\bar{\mu}^{n-1/2}/\Delta t + \bar{\kappa}^{n-1/2}/2}, \qquad \beta_{\mathrm{h}}^{n} = \frac{g_{\mathrm{h}}}{\bar{\mu}^{n}/\Delta t + \bar{\kappa}^{n}/2}, \tag{11}$$

where $\bar{\mu}$ is the effective permeability, $\bar{\kappa}$describes magnetic loss when included, and $g_{\mathrm{h}}$ is the dual-grid geometric factor. The numerical distinction between passive and dynamically switched regions was retained throughout the calculation. Material-dependent update coefficients in passive regions were calculated once before time stepping, whereas those associated with the active elements were updated whenever the prescribed digital state changed. This avoids repeatedly recomputing coefficients over the entire computational domain and confines the explicit time dependence to physically modulated regions.

The active-element representation was calibrated against the static EM response of the meta-atom used in this work. This preserved the numerical structure of the time-domain method while making the device model specific to the present metasurface. The incident field was monochromatic at $f_c$. The four time-dependent diode states were driven according to the same $[x_1x_2y_1y_2]$ sequences used experimentally. The calculation continued until the modulated response reached a periodically steady state. Time-domain fields were then Fourier transformed to obtain the complex amplitudes at $f_c + mf_s$. The time step obeyed the Courant stability condition, and absorbing boundaries surrounded the computational region. Static CST simulations and dynamic TVLE-TDFIT simulations therefore addressed complementary aspects of the experiment. CST provided the calibrated scattering alphabet and finite-aperture response for stationary configurations, whereas TVLE-TDFIT explicitly resolved temporal switching, frequency conversion and the phase-coherent formation of harmonic channels.

**Complex-field measurements and state-space projections**

The experimentally generated states were characterized through phase-coherent measurements of the EM field. Both orthogonal polarization components were referenced to the same excitation so that the relative phase between the two branches was preserved. This phase information is essential because the diagonal modal populations alone cannot distinguish the coherent Bell-like or GHZ-like states from incoherent mixtures containing the same populations. The measured complex field was projected onto the product basis defined by the active set $\mathcal{S}_q$. The projection operation depended on the physical state space. Frequency states were resolved through the corresponding harmonic components. Wavevector-direction states were separated in transverse spatial-frequency space. OAM states were evaluated through overlap with their prescribed azimuthal modes, and amplitude-envelope states were projected onto the corresponding normalized envelope basis. Polarization projection was performed in the orthogonal polarization basis shared throughout the work. For a spatial mode $u_\lambda(r)$, the corresponding complex modal coefficient of polarization component $p$ can be written as

$$a_{\lambda p} = \iint_\Omega u_\lambda^*(\boldsymbol{r}) E_p(\boldsymbol{r}) \mathrm{d}^2\boldsymbol{r}, \tag{12}$$

where $\Omega$ denotes the common analysis window. Equivalent projection operations were used in the spectral and wavevector-direction domains. All basis functions were normalized before state reconstruction, and identical projection definitions were applied to theory, full-wave simulation and experiment. For an arbitrary product-basis state $\mid b\rangle_{\mathcal{S}_q}$, the resulting coefficient is compactly written as

$$c_{\boldsymbol{b}}^{(q)} = {}_{\mathcal{S}_q}\left\langle \boldsymbol{b} \middle| \Psi_{\mathrm{meas}}^{(q)} \right\rangle. \tag{13}$$

The complete $D_q$-dimensional coefficient vector is then ordered according to the same binary product basis used in **Eq. 1**. Thus, the reconstructed state dimension was $D_q = 4, 8, 16$ and $32$ for the bipartite, tripartite, four-partite and five-partite demonstrations, respectively.

**Reconstruction of projected-state density matrices**

The experimentally reconstructed matrices in **Figs. 2-4** were obtained from the measured complex modal coefficients rather than from intensity data alone. For configuration $q$, the coefficient vector $\boldsymbol{c}_q = \left(c_1, c_2, \dots, c_{D_q}\right)^{\mathrm{T}}$ was normalized and converted into the projected-state coherency matrix

$$\rho_{\mathrm{meas}}^{(q)} = \frac{\boldsymbol{c}_q \boldsymbol{c}_q^\dagger}{\boldsymbol{c}_q^\dagger \boldsymbol{c}_q}. \tag{14}$$

For the deterministic coherent fields considered here, **Eq. 14** is a unit-trace rank-one coherency matrix by construction. Its diagonal entries resolve the projected modal populations, whereas its off-diagonal entries retain the relative phases of the measured complex modal coefficients. Within this coherent-field description, the complementary-branch entries identify the phase relation that is absent from an incoherent mixture with the same populations. Assessing statistical dephasing across repeated preparations instead requires the normalized ensemble operator $\langle \mathbf{aa}^{\dagger} \rangle / \mathrm{Tr}\langle \mathbf{aa}^{\dagger} \rangle$, rather than inferring ensemble purity from a single outer product (**Supplementary Note 13**). The matrix in **Eq. 14** is used as a classical coherency representation of the projected field. Density-matrix notation is adopted because the same tensor-product algebra is used to characterize the corresponding Bell-like and GHZ-like structures, but no quantum statistical interpretation, particle entanglement or Bell nonlocality is implied. The same reconstruction procedure was applied to theoretical, simulated and measured fields. No target-state phase was inserted during reconstruction. Relative phases entering the off-diagonal elements were obtained from the measured complex field coefficients themselves.

### Fidelity and element-wise state deviation

Agreement between the reconstructed and prescribed state was quantified using the same fidelity employed throughout the main text. For the ideal target $| \Psi_{q,\boldsymbol{b}}^{\pm} \rangle_{\mathrm{C}}$,

$$\mathcal{F}_q = {}_{\mathrm{C}}\langle \Psi_{q,\boldsymbol{b}}^{\pm} \left| \rho_{\mathrm{meas}}^{(q)} \right| \Psi_{q,\boldsymbol{b}}^{\pm} \rangle_{\mathrm{C}}. \tag{15}$$

This definition is used for all Bell-like and GHZ-like families in **Figs. 2-4** and therefore permits direct comparison across different physical compositions and multipartite orders. Because the projected states are normalized before evaluation, $\mathcal{F}_q$ characterizes agreement in state space rather than absolute scattering efficiency. The deviation maps were calculated element by element from

$$\Delta_{ij}^{(q)} = \left| \rho_{\mathrm{meas},ij}^{(q)} - \rho_{\mathrm{tar},ij}^{(q)} \right|, \tag{16}$$

with $\rho_{\mathrm{tar}}^{(q)} = \left| \Psi_{q,\boldsymbol{b}}^{\pm} \right\rangle_{\mathrm{CC}} \left\langle \Psi_{q,\boldsymbol{b}}^{\pm} \right|$. Identical basis ordering, normalization and spatial or spectral analysis windows were used for experimental and ideal matrices.

## Acknowledgements

This work was supported by the National Natural Science Foundation of China (62288101 and 92167202), the National Key Research and Development Program of China (2023YFB3813100 and 2022YFA1404903), the SEU Innovation Capability Enhancement Plan for Doctoral Students (CXJH_SEU 26043), Independent Research Fund of the State Key Laboratory of Millimeter Waves (Z202602-01), Jiangsu Joint Laboratory of Multidimensional Perceptual Information Technology (BM2022017), the State Key Laboratory of Millimeter Waves, Southeast University, China (K201924), the China Postdoctoral Science Foundation (2021M700761, 2022T150112), and the 111 Project (111-2-05).

## Author contributions

L.C., X.C., J.Y. and T.C. conceived the idea. L.C., Z.C., Q.Y., Z.H. and Z.Y. designed the experiments. L.C., J.Y. and T.C. supervised the project. L.C., Z.C., Q.Y. and Z.H. conducted the experiments, collected and analyzed the data. L.C. and Z.C. carried out the simulations, theoretical analyses, and wrote all the code. L.C., Z.L., X.C., J.Y. and T.C. wrote the manuscript, with contributions from all authors.

## Competing interests

The authors declare no competing interests.

## Data and materials availability

All data needed to evaluate the conclusions in the paper are present in the paper and the Supplementary Materials.

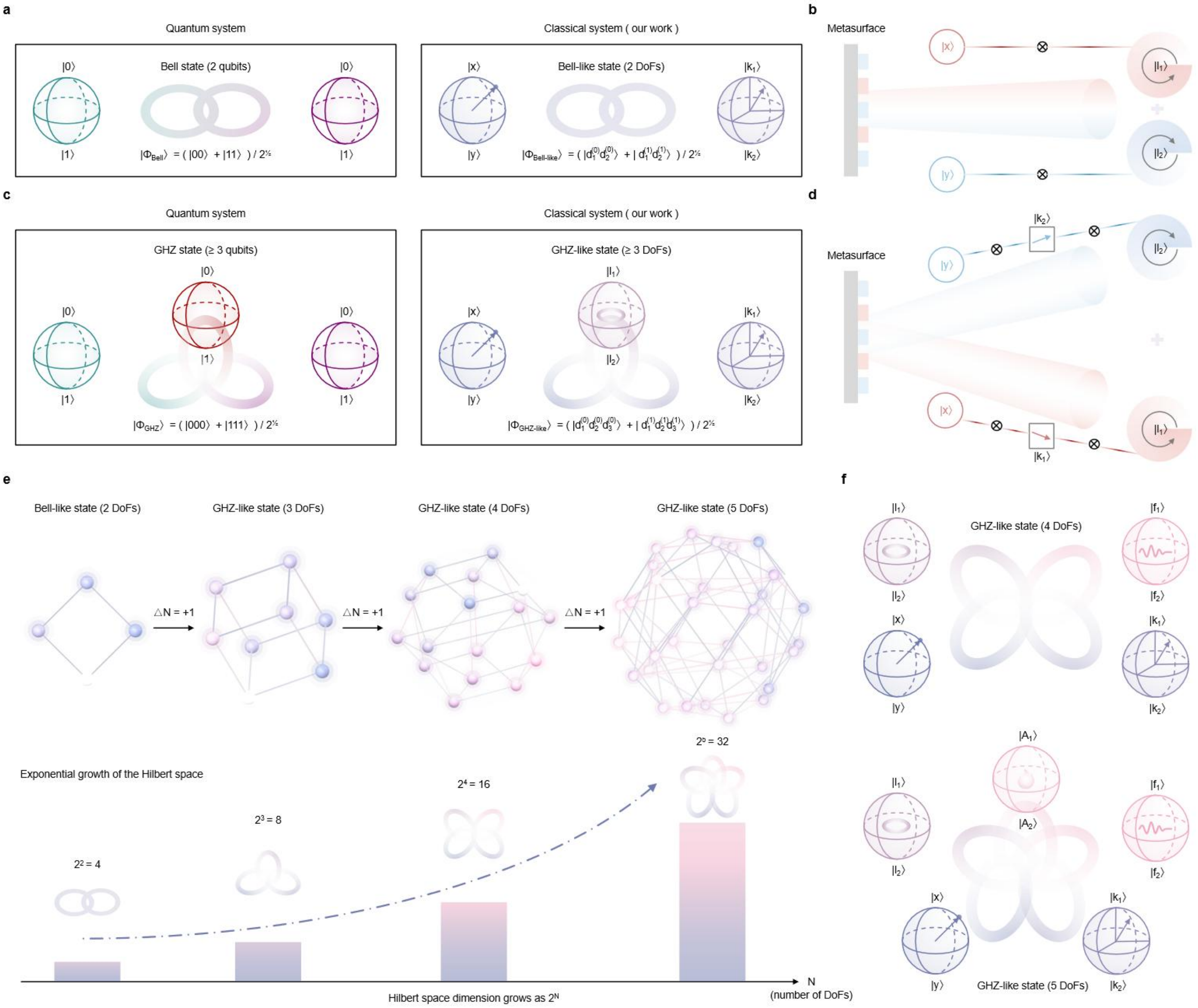

**Fig. 1 | State-space organization of programmable multipartiteness.** **a** Structural correspondence between a two-qubit Bell state and a classical Bell-like EM state. In the classical representation, two independently resolvable EM degrees of freedom replace the quantum subsystems while retaining the same nonfactorizable tensor-product structure. **b** Representative bipartite composition formed by polarization $P$ and OAM $L$. The two polarization basis states are coherently correlated with orthogonal OAM states, such that projection onto either polarization conditionally selects the corresponding OAM component. **c** Extension from bipartite to multipartite organization through the structural correspondence between a three-partite GHZ state and a classical GHZ-like EM state. The defining coherence connects two complementary product branches across all participating state spaces. **d** Representative tripartite EM composition formed by polarization $P$, wavevector $K$ and OAM $L$. **e** Hierarchy of binary multipartite state spaces from $N = 2$ to $N = 5$. The multipartite order $N$ counts the number of coherently participating state spaces, whereas the associated composite-space dimension follows $D = 2^N$, yielding $D = 4,8,16$ and 32. Dimensional growth is a consequence of increasing multipartite order and is distinct from enlarging the local dimension of a single degree of freedom. **f** Representative four- and five-partite compositions assembled from polarization $P$, frequency $F$, wavevector $K$, OAM $L$ and amplitude-envelope mode $A$. Changing the identity of the participating state spaces reconfigures the physical composition, whereas changing their number reconfigures the multipartite order. The five-partite configuration incorporates all five available EM state spaces into a common nonfactorizable structure.

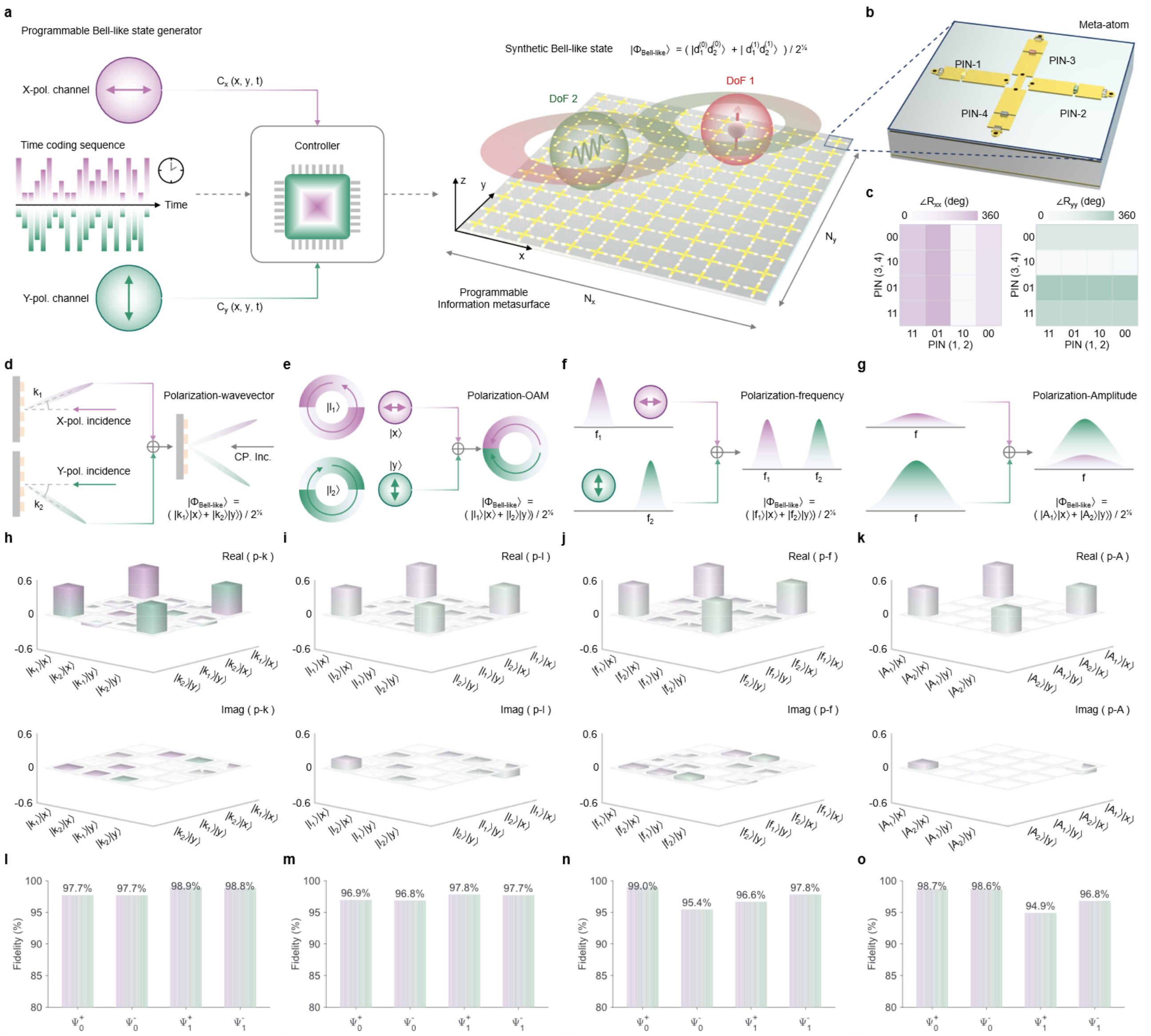

**Fig. 2 | Programmable state-space composition of Bell-like nonseparability. a** Schematic of the programmable bipartite-state generator. The two orthogonal polarization branches share the same physical aperture but are independently modulated by spatiotemporal coding functions $C_x(x, y, t)$ and $C_y(x, y, t)$, allowing each branch to be conditionally mapped onto a selected state of a second EM degree of freedom while the multipartite order remains $N_q = 2$. Magenta and green denote the $x$- and $y$-polarized branches, respectively. **b** Four-diode meta-atom comprising two independently addressable 2-bit phase-control channels for the orthogonal polarizations. **c** Relative reflection phases of the two polarization channels for their accessible 2-bit bias states. **d-g** Bell-like states with compositions $\mathcal{S}_q = \{P, K\}, \{P, L\}, \{P, F\}$, and $\{P, A\}$, respectively. Propagation-direction states $K$ are generated by spatial phase gradients, OAM states $L$ by azimuthal phase winding, frequency states $F$ by temporal modulation, and orthogonal amplitude-envelope states $A$ by spatial field shaping. In each case, the two polarization branches are coherently correlated with orthogonal states of the second factor. **h-k** Real and imaginary parts of the experimentally reconstructed projected-state density matrices for a representative $\Psi_0^+$ state in the four corresponding product spaces. **l-o** Measured fidelities of the complete Bell-like basis $\{\Psi_0^+, \Psi_0^-, \Psi_1^+, \Psi_1^-\}$ for each state-space composition.

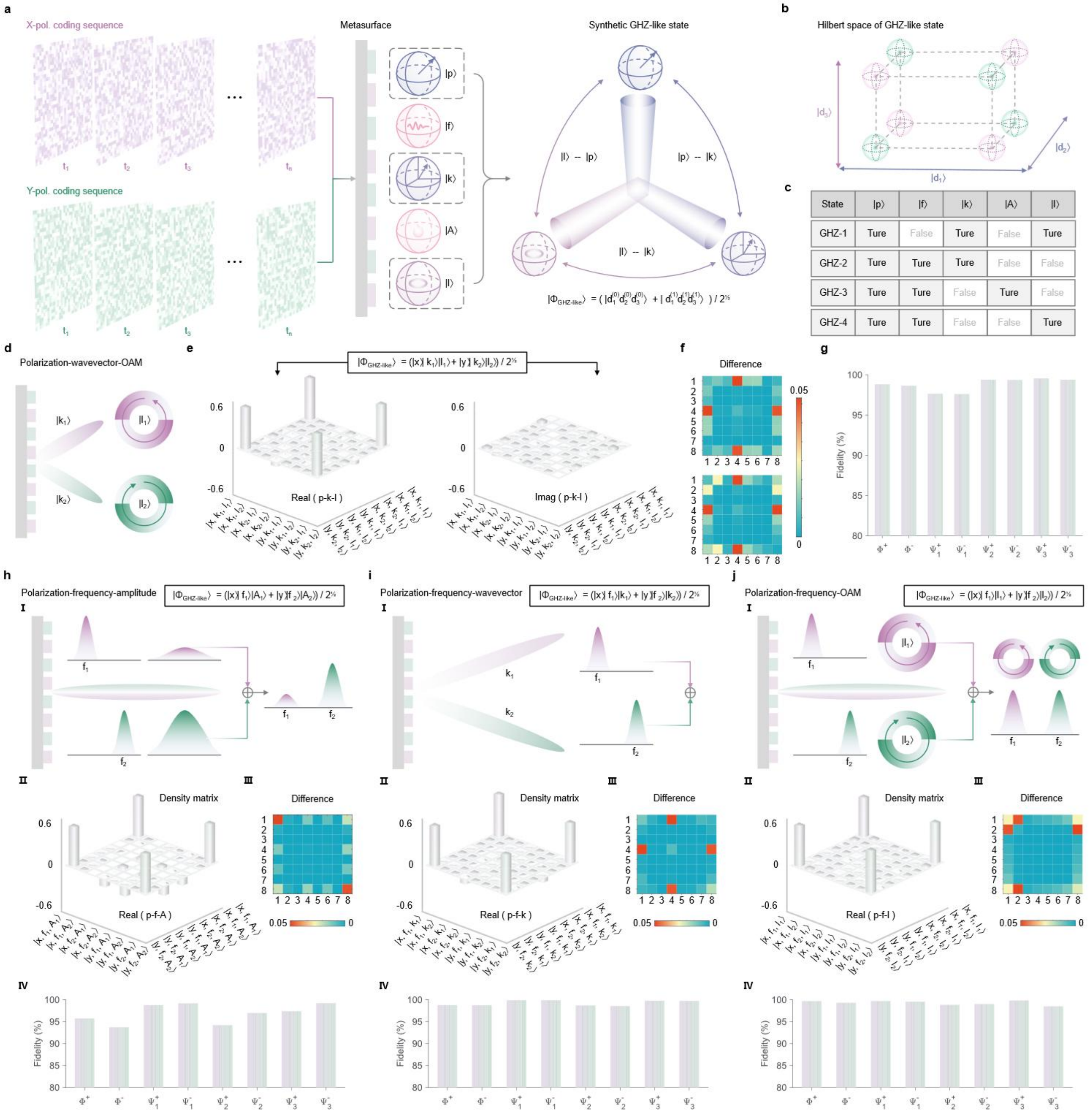

**Fig. 3 | Programmable global coherence across tripartite state spaces. a** Schematic of programmable tripartite GHZ-like state synthesis. Independent spatiotemporal coding of the two orthogonal polarization branches conditionally correlates polarization with two additional state-space factors selected from frequency $F$, wavevector $K$, OAM $L$ and orthogonal amplitude-envelope mode $A$. **b** Eight-dimensional product space for three binary factors, with the eight vertices representing the corresponding product-basis states at fixed multipartite order $N_q = 3$. **c** Participation map of four physically distinct tripartite compositions, $\{P, K, L\}$, $\{P, F, A\}$, $\{P, F, K\}$ and $\{P, F, L\}$, all with $N_q = 3$ and $D_q = 8$. **d-g** Characterization of the $P - K - L$ composition. Complementary three-factor branches forming the GHZ-like state (d). Real and imaginary parts of the reconstructed projected-state density matrix (e). Element-wise deviation of the reconstructed density matrix from the ideal target (f). Measured fidelities of the complete eight-state GHZ-like basis (g). **h-j** Corresponding realizations for the $P - F - A$ (h), $P - F - K$ (i) and $P - F - L$ (j) compositions, respectively. Each panel contains the state-space construction, reconstructed density matrix, deviation from the ideal target and fidelities of the complete GHZ-like basis.

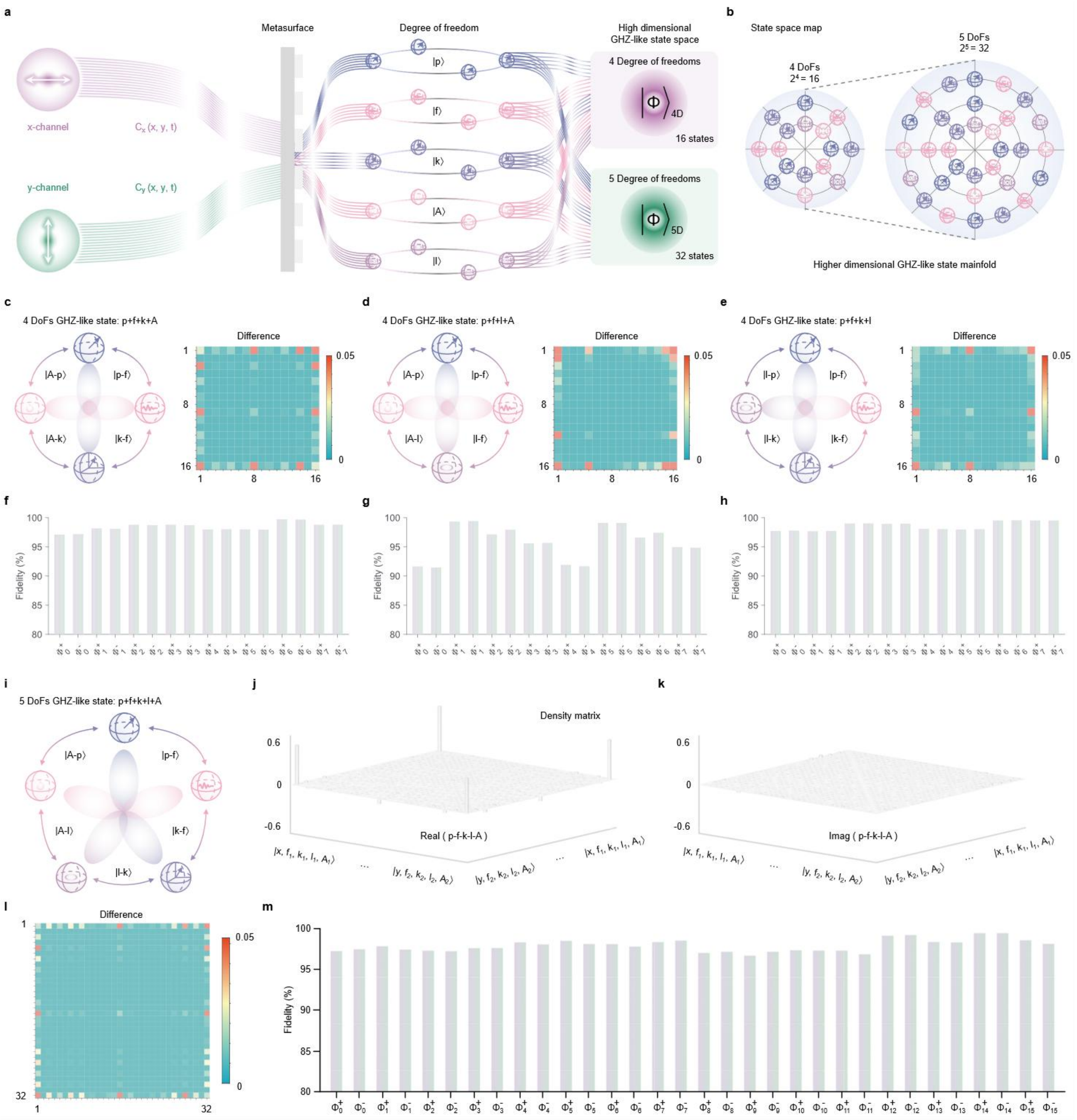

**Fig. 4 | Programmable scaling of GHZ-like coherence to five-partite order. a** Spatiotemporal programming of multipartite GHZ-like states using the same dual-polarization aperture. Different programming configurations select the identity and number of coherently participating state spaces from polarization $P$, frequency $F$, wavevector $K$, OAM $L$ and orthogonal amplitude-envelope mode $A$. Four active factors form a 16-dimensional composite space, whereas participation of all five factors produces a 32-dimensional space. **b** State-space hierarchy for $N_q = 4$ and $N_q = 5$, illustrating the extension of a common complementary-branch coherence across an increasing number of physical state spaces. **c-e** Four-partite GHZ-like compositions $P - F - K - A$ (c), $P - F - L - A$ (d) and $P - F - K - L$ (e), respectively. All three retain $N_q = 4$ and $D_q = 16$ while exchanging one physical constituent. The accompanying maps show element-wise deviations of the reconstructed projected-state density matrices from their ideal targets. **f-h** Measured fidelities of the complete 16-state GHZ-like families for the corresponding four-partite compositions. **i** Five-partite $P - F - K - L - A$ composition in which all available state spaces participate in the

same nonfactorizable branch structure. **j,k** Real and imaginary parts of the reconstructed projected-state density matrix for a representative five-partite GHZ-like state. **l** Element-wise deviation from the ideal target. **m** Measured fidelities of the complete 32-state five-partite GHZ-like family.